\documentclass[a4paper]{jpconf}
\usepackage{graphicx}
\usepackage{bm}
\usepackage{amsmath}
\newcommand{\la}{\langle}
\newcommand{\mbf}{\boldsymbol}
\newcommand{\ra}{\rangle}

\begin{document}
\title{Hydrogen atom as a quantum-classical hybrid system}

\author{Fei Zhan$^{1,2}$ and Biao Wu$^{1}$}

\address{$^1$ International Center for Quantum Materials, Peking University, 100871, Beijing, China

$^2$ Centre for Engineered Quantum Systems, School of Mathematics and Physics,
The University of Queensland, St Lucia QLD 4072, Australia}

\ead{wubiao@pku.edu.cn}

\begin{abstract}
Hydrogen atom is studied as a quantum-classical hybrid system, where the proton is treated
as a classical object while the electron is regarded as a quantum object. We use a well known
mean-field approach to describe this hybrid hydrogen atom; the resulting dynamics for 
the electron and the proton is compared to their  full quantum dynamics.  The electron 
dynamics in the hybrid description is found to be only marginally different from its
full quantum counterpart. The situation is very different for the proton: 
in the hybrid description, the proton behaves like a free particle;
in the fully quantum description, the wave packet center of the proton orbits around the
center of mass. Furthermore, we find that the failure to describe the proton dynamics 
properly can be regarded as a manifestation of the fact that there is no conservation 
of momentum in the mean-field hybrid approach. We expect that such a failure is a 
common feature for all existing approaches for quantum-classical hybrid 
systems of Born-Oppenheimer type. 
\end{abstract}

\section{Introduction}

Though quantum mechanics and its ensuing developments have been verified 
by all the experiments to be the correct description of the physical universe, 
classical physics is still a useful,  convenient, and sometimes irreplaceable  
tool to describe many systems. For example, it is very convenient and also 
accurate to describe the motion of the Earth around the Sun with Newton's
equations of motion instead of the Schr{\"o}dinger equation.  
There are even various situations, where one finds it convenient 
and necessary to separate one system into two subsystems: one subsystem
is described quantum mechanically and the other is described classically. 
This gives rise to quantum-classical hybrid systems. 

Quantum-classical hybrid systems come in three main categories: 
({\it i}) In unifying  gravity with the other fundamental interactions, 
one simply gives up quantizing gravity and proposes a unification theory where
gravity is treated as a classical field and the other three forces as quantum fields \cite{arlen_anderson1995prl,kuochungi1993prd,calzetta1994prd}. 
({\it ii}) In the standard Copenhagen interpretation of quantum mechanics, one always
deals with a hybrid system, where a quantum system interacts with a classical measuring 
apparatus. In such a case, the classical apparatus can cause the collapse of the wave function
of the quantum system, which is still beyond mathematical description \cite{zurek2003rmp}. 
({\it iii}) The hybrid system in this category is typified by systems in the field of
solid state physics and chemistry \cite{tully1971jcp,tully1990jcp,prezhdo1997jcp}. In the Born-Oppenheimer
approximation \cite{born1927adp},  the  nuclei are treated classically  because their motion 
is much slower due to their large masses while the electrons are treated as quantum objects. 
This kind of systems are now also arising in nano-science \cite{rugar2004n,zhangqi2006prl}, where
a classical detector of tens of nanometers in size interacts with a quantum object. 
For these systems, the interaction from the classical subsystem does not cause dramatic
changes in the quantum subsystem such as the collapse of the wave function. 
For systems in this category, we call them Born-Oppenheimer systems.

We are only interested in the Born-Oppenheimer hybrid systems. There have been many 
different approaches proposed to describe these hybrid systems\cite{tully1971jcp,tully1990jcp,jones,prezhdo1997jcp,Aleksandrov,Beswick,Delos1972PRA,Delos1972pra2,Halliwell,Jones,Kapral1999,Nielsen2001,Prezhdo3,Rossky,castro}; some of these approaches are 
shown to be equivalent to each other \cite{zhanfei2008jcp}. There are 
discussions on what conditions need to be satisfied by a self-consistent theory 
for a hybrid system\cite{Caro,Diosi,elze2011jpcs,elze2012pra,Salcedo}.  In this work, we use hydrogen atom as a 
concrete example to explore the possible inconsistency or inadequacy of a theory 
for hybrid systems. As is well known,  the full quantum mechanical description 
of hydrogen atom, the simplest  Born-Oppenheimer system in nature, can be found 
analytically. This full quantum description can be used to benchmark the hybrid 
description. 

We use the mean-field approach that was proposed in Ref.~\cite{zhangqi2006prl} 
to describe hydrogen atom. Its theoretical structure was fully analyzed in Ref.~\cite{elze2012pra}. 
With this approach, we find that the electron dynamics is very similar to the corresponding 
full quantum description. However, we find that the proton in the hybrid approach 
behaves like a free particle, very different from its full quantum description where
the wave packet center of the proton orbits around the center of mass. This failure 
to describe the proton motion properly is rooted in the fact 
that there is no conservation of momentum in the hybrid approach.  Our further
analysis shows that this failure or inadequacy in the hybrid approach is intrinsic:
it is caused by the loss of entanglement of the electron and proton dynamics 
in the mean-field hybrid approach. As this loss of entanglement  exists in all
known hybrid approaches, we expect that all the hybrid approaches 
fail to describe properly the dynamics of the classical subsystem. 



\section{Full quantum solution for hydrogen atom}
\label{section:coarsegrain}
Before we study the hydrogen atom as a hybrid system, we briefly review 
its general full quantum solution  and then apply it to a special case, where
the electron is described by a wave packet that moves and evolves around
a circle \cite{gaeta1990pra}. 

The Schr\"odinger equation for a hydrogen atom can be written as
\begin{equation}\label{eqn:schrodinger1}
 i\hbar\frac{\partial}{\partial t}\psi(\mbf{r}_{e},\mbf{r}_{p},t)=
 \left[-\frac{\hbar^2}{2m_{e}}\nabla^2_{\mbf{r}_{e}}-\frac{\hbar^2}{2m_{p}}
 \nabla^2_{\mbf{r}_{p}}+V(|\mbf{r}_{e}-\mbf{r}_{p}|)
 \right]\psi(\mbf{r}_{e},\mbf{r}_{p},t),
\end{equation}
where $m_p$ ($\mbf{r}_{p}$) is the mass (coordinate) of the proton, 
$m_e$ ($\mbf{r}_{e}$) is the mass (coordinate) of the electron, and 
$V(|\mbf{r}_{e}-\mbf{r}_{p}|)$ is the Coulomb potential that depends 
only on  the distance between the proton and the electron. As done in every textbook, 
we introduce  the relative coordinate $\mbf{r}$ and the coordinate of the center
of mass $\mbf{R}$, 
\begin{equation}
 \mbf{r}=\mbf{r}_{e}-\mbf{r}_{p},~~\mbf{R}=\frac{m_{e}\mbf{r}_{e}+m_{p}\mbf{r}_{
p}}{M},\label{coordinates_transform}
\end{equation}
where $M=m_{e}+m_{p}$ is the total mass of the hydrogen atom.  In these new 
coordinates, the Schr\"odinger equation \eqref{eqn:schrodinger1} becomes
\begin{equation}\label{eqn:schrodinger2}
 i\hbar\frac{\partial}{\partial
t}\psi(\mbf{R},\mbf{r},t)=\left[-\frac{\hbar^2}{2M}\nabla^2_{\mbf{R}}-\frac{\hbar^2}{2\mu}\nabla^2_{
\mbf{r}}+V(|\mbf{r}|)\right]\psi(\mbf{R},\mbf{r},t),
\end{equation}
where $\mu=m_{e}m_{p}/(m_{e}+m_{p})$ is the reduced mass. So, the wave function
of a hydrogen atom can be written as
\begin{equation}
\label{sep}
\psi(\mbf{R},\mbf{r},t)=\psi_{r}(\mbf{r},t)\psi_{c}(\mbf{R},t)\,,
\end{equation}
where $\psi_{r}(\mbf{r},t)$ and $\psi_{c}(\mbf{R},t)$ satisfy, respectively, 
\begin{align}
 i\hbar\frac{\partial}{\partial
t}\psi_r(\mbf{r},t)&=-\frac{\hbar^2}{2\mu}\nabla^2_{\mbf{r}}\psi_{r}(\mbf{r},t)+
V(|\mbf{r}|)\psi_{r}(\mbf{r},t)\label{relativeeqn}\,,\\
i\hbar\frac{\partial}{\partial
t}\psi_c(\mbf{R},t)&=-\frac{\hbar^2}{2M}\nabla^2_{\mbf{R}}\psi_c(\mbf{R},t)\,.\label{centermasseqn}
\end{align}
This shows that the motion of a hydrogen atom can be separated
into two independent parts: $\psi_r(\mbf{r},t)$ describes the motion of a particle
in a Coulomb potential $V(|\mbf{r}|)$ while $\psi_c(\mbf{R},t)$ describes the motion of
a free particle. However, this by no means implies that the motions of the electron
and the proton can be described by two independent wave functions, which will
become much clearer in the later analysis.  

We go back to the coordinate system of $\mbf{r}_{e}$ and $\mbf{r}_{p}$. 
The density distribution of the electron is
\begin{align}
|\psi_{e}(\mbf{r}_{e},t)|^2&=\int |\psi_{r}(\mbf{r},t)|^2|\psi_{c}(\mbf{R},t)|^2d\mbf{r}_{p}\notag\\
&=\int |\psi_{r}(\mbf{r}_e-\mbf{r}_p,t)|^2|\psi_{c}(\frac{m_{e}\mbf{r}_{e}+m_{p}\mbf{r}_{
p}}{M},t)|^2d\mbf{r}_{p}\notag\\
&=\int |\psi_{r}(\mbf{r}_e-\mbf{r}_p,t)|^2|\psi_{c}(\mbf{r}_{e}-\frac{m_{p}}{M}(\mbf{r}_{e}-\mbf{r}_{p}),t)|^2d(\mbf{r}_{p}-\mbf{r}_{e})\notag\\
&=\int |\psi_{r}(\mbf{x},t)|^2|\psi_{c}(\mbf{r}_{e}-\frac{m_{p}}{M}\mbf{x},t)|^2d\mbf{x}\,,
\label{cgelectron}
\end{align}
where we have plugged in Eq. \eqref{coordinates_transform}.  Similarly, the  proton density is
given by
\begin{equation}
\label{cgproton}
|\psi_{e}(\mbf{r}_{p},t)|^2=\int |\psi_{r}(\mbf{x},t)|^2|\psi_{c}(\mbf{r}_{p}-\frac{m_{e}}{M}\mbf{x},t)|^2d\mbf{x}\,.
\end{equation}
The above results are very illuminating. As the center of mass motion is free particle-like, 
we assume that $\psi_{c}(\mbf{R},t)$ is a Gaussian wave packet with width $\sigma$. 
So, the density of electron is just the density of the relative motion $|\psi_{r}(\mbf{r},t)|^2$
coarse-grained with a Gaussian function of width $M\sigma/m_p\approx \sigma$.
In contrast, the density of proton is the density of the relative motion $|\psi_{r}(\mbf{r},t)|^2$
coarse-grained with a Gaussian function of a much larger width $M\sigma/m_e\approx 1837\sigma$.
In other words, the electron density and the proton density are very similar to each other but
the proton density looks about 1837 times fuzzier. 

In the reference frame where the hydrogen atom is motionless, the center of the wave packet
of electron is
\begin{eqnarray}
\label{electronc}
\la \mbf{r}_{e}\ra=\frac{m_p}{M}
\int|\psi_{r}(\mbf{r},t)|^2\mbf{r}d\mbf{r}=\frac{m_p}{M}\la \mbf{r}\ra\,,
\end{eqnarray}
where $\la \mbf{r}\ra$ is the center of the wave packet for the relative motion.
The center of the wave packet of proton is
\begin{eqnarray}
\label{protonc}
\la \mbf{r}_{p}\ra=-\frac{m_e}{M}\la \mbf{r}\ra\,.
\end{eqnarray}

We now consider a special case. In this case, the initial state of 
the center of mass of the hydrogen atom is described by a  Gaussian function 
\begin{equation}
\psi_c(\mbf{R},t=0)=\frac{1}{(2\pi\sigma^2)^{1/4}}
\exp\left[-\frac{|\mbf{R}|^2}{4\sigma^2}\right]
\end{equation}
with $\sigma$ being the width; the relative motion at the beginning is depicted 
by the following wave function \cite{gaeta1990pra},
\begin{equation}
\label{circle}
\psi_r(\mbf{r},t=0)=\frac{1}{(2\pi\sigma_{\bar{n}}^2)^{1/4}}\sum_{n=1}^{\infty}\exp\left[-\frac{(n-\bar{n})^2}{4\sigma_{\bar{n}}^2}\right]u_{n(n-1)(n-1)}(\mbf{r})\,,
\end{equation}
where $\bar{n}$ and $\sigma_{\bar{n}}$ are the mean and the width of the Gaussian distribution, respectively, 
and $u_{nlm}$ is the standard energy-eigenstate for the hydrogen atom\cite{schiff1955}
\begin{equation}\label{unlm}
 u_{nlm}=\sqrt{\left(\frac{2}{na_B}\right)^{3}\frac{(n-l-1)!}{2n[(n+1)!]^{3}}}e^{-r/na_B}\left(\frac{2r}{na_B}\right)^{l}L_{n-l-1}^{2l+1}\left(\frac{2r}{na_B}\right)Y_{l}^{m}(\theta,\phi)
\end{equation}
with $L$ being associated Laguerre polynomial, $Y$ the spherical harmonics, and
$a_B$ the Bohr radius.  

The ensuing dynamics of the circular wave packet in Eq.(\ref{circle}) under the influence of
the Coulomb potential $V(|\mbf{r}|)$ has been studied in detail in Ref.\cite{gaeta1990pra}. 
For the sake of self-containment, we summarize their results here.  The wave packet is
localized on a circle with radius $\sim\bar{n}^2a_B$.  
This wave packet remains localized and moves on the circle for several $T_{\rm Kepler}$, 
the period of the corresponding classical motion on the same circle. At time  
$T_{\rm spread}\sim 10T_{\rm Kepler}$, the spreading of the wave packet becomes
so severe that it distributes rather uniformly on the circle. 
This is characterized by $\la \mbf{r}\ra=0$. The wave packet can recover its localized
form and revive at $T_{\rm rev}=(\bar{n}/3)T_{\rm Kepler}$, and repeats its previous
dynamics afterwards. With this in mind, it is straightforward to picture the 
quantum dynamic motion for both the electron and the proton in this special case.

For the electron,  its wave packet is the wave packet in Eq.(\ref{circle}) 
coarse-grained with the Gaussian wave packet for the center of mass 
motion (see Eq.(\ref{cgelectron})). As long as the width $\sigma$ is not
too large,  its wave packet dynamics should be very similar: 
it is localized and orbits on a circle of radius $\sim\bar{n}^2a_B$ before $T_{\rm spread}$. 
During this period,  the wave packet center $\la \mbf{r}_{e}\ra$ oscillates periodically
with its amplitude decreasing.  After $T_{\rm spread}$ and before $T_{\rm rev}$,  
the wave packet spreads over the circle, which is characterized by $\la \mbf{r}_{e}\ra=0$.  
This dynamics repeats itself after $T_{\rm rev}$. 

For the proton, its wave packet center has a similar motion as the electron's according
to Eqs.(\ref{electronc},\ref{protonc}). The difference is that the proton moves 
in the opposite direction and $\la \mbf{r}_{p}\ra$ varies with time on a circle 
of much smaller radius. However,  the dynamics of the proton wave packet 
is very different. The wave packet of the proton is the 
result of coarse-graining with a Gaussian function of 
large width $M\sigma/m_e\approx 1837\sigma$ (see Eq.(\ref{cgproton})). 
This large-size coarse-graining makes the wave packet much less localized. At the same time, 
the proton moves on a much smaller circular orbit (about 1837 times smaller). 
With these two factors combined, it is clear that the wave packet of the proton is 
always spread and smeared out over the entire circular orbit. No distinct peak and other
structure can be seen. The three different time scales of the wave packet dynamics, $T_{\rm Kepler}$, $T_{\rm spread}$,  and $T_{\rm rev}$, which are used to characterize
the distinct features of the wave packet at different times,   become rather meaningless for the proton.

\section{Hybrid dynamics in hydrogen atom}\label{section:hybr}
We now treat the hydrogen atom as a hybrid  system, where the proton
is regarded as a classical object while the electron is treated quantum mechanically.
With the approach in Ref.\cite{zhangqi2006prl}, the hybrid
Hamiltonian for the hydrogen atom is 
\begin{equation}\label{hamiltonian1}
H=\la\varphi_{e}(\mbf{r}_{e},t)|-\frac{\hbar^2}{2m_{e}}\nabla^2_{\mbf{r}_{e}}+
V(|\mbf{r}_{e}-\mbf{r}_{p}|)|\varphi_{e}(\mbf{r}_{e},t)\ra+\frac{\mbf{p}_{p}^2}{2m_{p}}.
\end{equation}
We expand the wave function $|\varphi_{e}\ra$ in a set of complete orthonormal basis,
\begin{equation}
\label{exp}
|\varphi_{e}(\mbf{r}_{e},t)\ra=\sum_{j}\varphi_{j}(t)|j\ra. 
\end{equation}
With the classical canonical Hamiltonian structure introduced by Heslot\cite{heslot1985prd}, 
we have the following Poisson brackets,
\begin{align}
&\{\varphi_{j},\varphi_{k}^{*}\}=i\delta_{jk}/\hbar,~~\{r_{pj},p_{pk}\}=\delta_{jk},\\
&\{\varphi_{j},\varphi_{k}\}=\{r_{pj},r_{pk}\}=\{p_{pj},p_{pk}\}=0,
\end{align}
where  $r_{pj}(p_{pj})$ is the $j$th component of coordinate (momentum) vector of the proton.
With these Poisson brackets, we can obtain the hybrid equations of motion, 
\begin{align}
 i\hbar\frac{d}{dt}|\varphi_{e}(\mbf{r}_{e}, t)\ra&=\left[-\frac{\hbar^2}{2m_{e}}\nabla^2_{\mbf{r}_{e}}+V(\mbf{r}_{e}-\mbf{r}_{p} )\right ] |\varphi_{e}(\mbf{r}_{e},t)\ra\,,
 \label{electrond}\\
\dot{\mbf{r}}_{p}&=\frac{\partial H}{\partial \mbf{p}_{p}}=\frac{\mbf{p}_{p}}{m_{p}}\,,
\label{protonp1}\\
\dot{\mbf{p}}_{p}&=-\frac{\partial H}{\partial
\mbf{r}_{p}}=-\nabla_{\mbf{r}_{p}}\la\varphi_{e}(\mbf{r}_{e},t)|-\frac{\hbar^2}{2m_{e}}\nabla^2_{\mbf{r}_{e}}+V(\mbf{r}_{e}-\mbf{r}_{p} )|\varphi_{e}(\mbf{r}_{e},t)\ra\,.
\label{protonp}
\end{align}
Note that $\mbf{p}_{p}$ and $\mbf{r}_{p}$ are independent dynamical variables
in the above equations of motion while $\mbf{r}_{e}$ is just some external parameter. 
The dynamical variables for the electrons are $\varphi_j$'s in Eq.(\ref{exp}). 

Similar to the full quantum treatment, we focus on the case where
the initial state for the electron  is given by the circular wave packet in Eq.(\ref{circle}). 
In terms of the electron coordinate $\mbf{r}_e$, the circular wave packet has the following form,
\begin{equation}
\label{ecircle}
\varphi_e(\mbf{r}_e,t=0)=\frac{1}{(2\pi\sigma_{\bar{n}}^2)^{1/4}}\sum_{n=1}^{\infty}\exp\big[-\frac{(n-\bar{n})^2}{4\sigma_{\bar{n}}^2}\big]u_{n(n-1)(n-1)}(\mbf
{r} _e-\mbf{r}_{p0})\,,
\end{equation}
where $\mbf{r}_{p0}$ is the initial position of the proton. Since the proton is much
more massive than the electron, its motion is rather slow and will not cause quantum transition
between different quantum states of the electron. In other words, the weight before 
each eigenstate $u_{n(n-1)(n-1)}$ will not be changed by the proton motion. This is
just the famous quantum adiabatic theorem\cite{born1928zp} or the essence of the Born-Oppenheimer approximation. 
As a result, the dynamics of the electron wave packet is just the same as the electron dynamics
in the full quantum description. The only difference is that the wave function of the electron 
in this hybrid approach is not coarse-grained. 

In contrast, the proton is very different: the classical motion of a proton in the hybrid treatment
does not agree with the motion of the wave packet center $\la \mbf{r}_{p}\ra$ for any meaningful
period of time.  Let us examine the right hand side of Eq.(\ref{protonp}). 
$E_e=\la\varphi_{e}|-\frac{\hbar^2}{2m_{e}}\nabla^2_{\mbf{r}_{e}}+
V(\mbf{r}_{e}-\mbf{r}_{p} )|\varphi_{e}\ra$ is the electron energy.  It is clear that
the eigen-energy $E_{nlm}$ for each eigenstate $u_{nlm}$ is independent of the 
proton position $\mbf{r}_{p}$. As mentioned above, the proton motion is so slow that 
it will not cause quantum transition between different electronic states $u_{nlm}$. This 
means that $E_e$ is independent of $\mbf{r}_{p}$. As a result, 
the right hand side of Eq.(\ref{protonp}) is zero and the momentum of the proton
does not change with time. So, the proton in the hybrid dynamics is  like a free-particle, 
motionless or moves along a straight line. In contrast, the wave packet center $\la \mbf{r}_{p}\ra$
makes circular motion around the center of mass before time $T_{\rm spread}$. 
This shows that the hybrid approach fails to describe the proton dynamics properly. 

This failure to describe the proton dynamics can be  illustrated from a different angle. 
 Our analysis shows that  this failure 
essentially has the root in the fact that there is no conservation of momentum 
in the hybrid dynamics of hydrogen atom.  We define the total momentum
for this system as
\begin{equation}
\mbf{P}=\mbf{p}_p+\la\varphi_{e}|\hat{\mbf{p}}_e|\varphi_{e}\ra\,.
\end{equation}
In the adiabatic limit, we know that $\mbf{p}_p$ is a constant while 
$\la\varphi_{e}|\hat{\mbf{p}}_e|\varphi_{e}\ra$ changes with time significantly 
at least for the first several $T_{\rm Kepler}$. This means that the total momentum $\mbf{P}$
changes with time. The easiest way to appreciate this result to set  
the initial momentum $\mbf{p}_{p0}$ of the proton be zero. Then according to 
Eqs.(\ref{electrond},\ref{protonp1},\ref{protonp}), the proton remains motionless
while the electron wave packet moves on the  circle and its change of momentum 
$d\la\varphi_{e}|\hat{\mbf{p}}_e|\varphi_{e}\ra/dt$ has  a finite value 
for the first several $T_{\rm Kepler}$.

\section{Self-consistency of hybrid approaches}
The above discussion leads to immediate questions, such as,  ``How general is the conclusion
that the hybrid approach is inadequate in describing the dynamics of the classical subsystem?", 
``Is it possible that hybrid approaches are intrinsically flawed?'' There is already some
attempt to answer these questions~\cite{Salcedo}. Again we do not do general analysis here; 
we instead use hydrogen atom as an example and hope that our analysis with hydrogen
atom may shed some light on the general questions. 

First, the analysis on hydrogen atom can be generalized to multi-nucleus systems. 
The electronic states clearly do not depend on the position of the center of mass of all
the nuclei. When the nuclear center of mass moves slowly as in the usual case, 
it will not cause quantum transition between different electronic states. As a result, 
the center of mass of all the nuclei feels no force and there is no conservation of the
total momentum.  It seems that this shortcoming can be remedied by moving
the derivative in  Eq.(\ref{protonp}) inside the bra-ket, that is, replacing Eq.(\ref{protonp})
with 
\begin{equation}
\dot{\mbf{p}}_{p}=-\frac{\partial H}{\partial
\mbf{r}_{p}}=-\la\varphi_{e}(\mbf{r}_{e},t)|\nabla_{\mbf{r}_{p}}\left[
-\frac{\hbar^2}{2m_{e}}\nabla^2_{\mbf{r}_{e}}+V(\mbf{r}_{e}-\mbf{r}_{p} )\right]
|\varphi_{e}(\mbf{r}_{e},t)\ra\,.
\label{protonp2}
\end{equation}
This is nothing but the well-known Ehrenfest equation\cite{alonso,Bornemann}. It is easy
to show that the total momentum is conserved if this  Ehrenfest equation is used instead 
of Eq.(\ref{protonp}).  As the Ehrenfest equation is derived from Eq.(\ref{protonp}) 
with an argument that appears right but is fundamentally flawed upon close 
examination\cite{Bornemann}, we now have a very 
intriguing situation:  the total momentum is not conserved for the ``correct" Eq.(\ref{protonp})
while the total momentum is conserved for the ``flawed" Eq.(\ref{protonp2}). 
This  may be regarded as an indication that there exists intrinsic inconsistency in
a hybrid theory. 

Secondly, there is loss of quantum entanglement in any hybrid theory. 
In the full classical treatment,  a pair of  independent dynamical variables 
$\{\mbf{r}_e(t),~\mbf{r}_p(t)\}$, one for the electron and the other for the proton, 
are enough to specify completely the dynamics of the system. In the hybrid approach, 
there also exists such a dynamical pair $\{\mbf{r}_p(t),\varphi_{e}(\mbf{r}_e, t)\}$, 
for the proton and  the electron, respectively, which can completely describe the
whole dynamics. 

In contrast, in the full quantum mechanical approach, there exists no
such a dynamical pair, one for the electron and the other for the proton,
with which the full quantum dynamics is completely determined. It is true according to
Eq.(\ref{sep}) that the full quantum dynamics can be specified by two independent
wave functions. However, one wave function is for the relative motion and the other
is for the center of mass, instead of for the electron and the proton, respectively. 
The motion of the electron and the proton is 
always entangled together.  This entanglement between the electron and the proton
is manifested at two levels. For the first level, it is the entanglement 
understood in the usual sense that one can not have the total wave function as the product of
the electron wave function and the proton wave function, $\psi(t)=\psi_e(\mbf{r}_e, t)\psi_p(\mbf{r}_p, t)$. Maybe temporarily at a given moment,  one can have this kind of direct 
product but not for any meaningful period of time.  For the second level, which is
much weaker, the total wave function can not be the functional of two independent 
dynamical functions, one for the electron $f_e(\mbf{r}_e, t)$ and the other 
for the proton $f_p(\mbf{r}_p, t)$. In other words, we can never write
the total wave function as $\psi(t)=\psi[f_e(\mbf{r}_e, t), f_p(\mbf{r}_p, t)]$.
To see this clearly,  we introduce the reduced wave functions for the electron
and the proton. Similar to the densities, they are also coarse-grained from
the wave function for the relative motion,  
\begin{equation}
\widetilde{\psi}_{e}(\mbf{r}_{e},t)=
\int \psi_{r}(\mbf{r},t)\psi_{c}(\mbf{R},t)d\mbf{r}_{p}=
\int \psi_{r}(\mbf{x},t)\psi_{c}(\mbf{r}_{e}-\frac{m_{p}}{M}\mbf{x},t)d\mbf{x}\,,
\label{cgelectronf}
\end{equation}
\begin{equation}
\widetilde{\psi}_{p}(\mbf{r}_{p},t)=
\int \psi_{r}(\mbf{r},t)\psi_{c}(\mbf{R},t)d\mbf{r}_{e}=
\int \psi_{r}(\mbf{x},t)\psi_{c}(\mbf{r}_{p}-\frac{m_{e}}{M}\mbf{x},t)d\mbf{x}\,.
\label{cgprotonf}
\end{equation}
It is clear that these two functions contain all the dynamical information that we can 
have for the electron and the proton. However, as they are obtained from the coarse-graining, 
these two reduced wave functions alone can not specify the total wave function $\psi(t)$. 
The entanglement between the electron and the proton at the second level is immediately destroyed 
in the hybrid approach, where the dynamics of the whole system can be determined completely
by the proton dynamics $\mbf{r}_{p}$ and the electron dynamics $\varphi_{e}(\mbf{r}_e, t)$.  
Although our analysis is done for hydrogen atom,  it can 
be generalized. In any hybrid approach, there is no entanglement between the quantum
subsystem and the classical subsystem at the second level. 
This loss of entanglement is general and intrinsic.  

\section{Conclusion}
\label{section:conclusion}
In summary, we have studied the dynamics in a hydrogen atom with two different approaches. 
One is the full quantum theory that can be found in the standard textbook and the other
is the hybrid mean-field approach, where the proton is regarded as a classical object 
and the electron is a quantum object. We have found that there is only marginal 
difference for the dynamics of the electron between these two approaches. However,
the dynamics for the proton is very different between the two: the proton in the hybrid
approach behaves like a free particle, not moving or moving along a straight line; 
the wave packet center for the proton
in the full quantum approach can make circular motion for a limited time. These
differences are summarized in Table 1. 

\begin{table}[!h]
\begin{tabular}{|l||c|c|}
\hline
 & full quantum approach & hybrid approach\\
 \hline
 \hline
 electron dynamics & coarse-grained $\psi_{r}(\mbf{r}, t)$ & $\psi_{r}(\mbf{r}, t)$\\
 \hline
 proton dynamics & $\la \mbf{r}_{p}\ra$ moves around the center of mass &
 behaves as a free particle\\
 \hline
 & conservation of the total momentum & no conservation of the total momentum\\
 \hline
\end{tabular}
\caption{Comparison between the full quantum dynamics and the hybrid dynamics of hydrogen atom}
\end{table}

The failure of proper description of the proton dynamics 
is the manifestation that there is no conservation of total momentum in the hybrid approach. 
We acknowledge that this failure may just be to the interest of theorists as there is
no experimental way to probe the proton dynamics. For systems with more than one nuclei, 
the non-physical artifacts caused by the hybrid treatment may also be not essential to experiments. 
The reason is that the hybrid approach may only fail to capture the proper motion for the center
of mass of the nuclei. For the  relative motions between the nuclei, 
the hybrid approach may be adequate.    
All the vibrational and rotational frequencies for the relative nuclear motion are 
computed by treating the nuclei as classical 
objects. If this approximation can not capture physics that can be measured experimentally,
it would be found and known for a long time.

\ack 
We thank Yinhan Zhang and Shaoqi Zhu for useful discussions. B.W. is supported by the National Basic Research Program of MOST (NBRP) of China (2012CB921300, 2013CB921900), the National Natural Science Foundation (NSF) of China (11274024), the Research Fund for the Doctoral Program of Higher Education (RFDP) of China (20110001110091). 
\section*{References}

\end{document}